\documentclass[conference]{IEEEtran}
   \usepackage[pdftex]{graphicx}
\usepackage{amsmath}
\DeclareMathSizes{10}{10}{8}{7}   

\begin{document}
%
\title{Random Linear Network Coding For Time Division Duplexing: Energy Analysis}


\author{\authorblockN{Daniel E. Lucani}
\authorblockA{RLE, MIT\\
Cambridge, Massachusetts, 02139\\
Email: dlucani@mit.edu}
\and
\authorblockN{Milica Stojanovic}
\authorblockA{Northeastern University\\
Boston, Massachusetts, 02118\\
Email: millitsa@mit.edu}
\and
\authorblockN{Muriel M\'edard}
\authorblockA{RLE, MIT\\
Cambridge, Massachusetts, 02139\\
Email: medard@mit.edu}}


%


\maketitle

\begin{abstract}
We study the energy performance of random linear network coding for time division duplexing channels. We assume a packet erasure channel with nodes that cannot transmit and receive information simultaneously. The sender transmits coded data packets back-to-back before stopping to wait for the receiver to acknowledge the number of degrees of freedom, if any, that are required to decode correctly the information. Our analysis shows that, in terms of mean energy consumed, there is an optimal number of coded data packets to send before stopping to listen. This number depends on the energy needed to transmit each coded packet and the acknowledgment (ACK), probabilities of packet and ACK erasure, and the number of degrees of freedom that the receiver requires to decode the data. We show that its energy performance is superior to that of a full-duplex system. We also study the performance of our scheme when the number of coded packets is chosen to minimize the mean time to complete transmission as in \cite{lucaniInfocom09}. Energy performance under this optimization criterion is found to be close to optimal, thus providing a good trade-off between energy and time required to complete transmissions. 
\end{abstract}


%
\IEEEpeerreviewmaketitle

\section{Introduction}

The concept of network coding, also known as coded packet networks, was introduced by Ahlswede \textit{et al} \cite{ahslwede00}. Network coding considers nodes that have a set of functions that operate upon received or generated data packets. A classical network's task is to transport packets provided by the source nodes unmodified, i.e. they constitute a subset of the coded packet networks, in which each node has two main functions: forwarding and replicating a packet. In contrast, network coding considers information as an algebraic entity, on which one can operate. 

Network coding research originally studied throughput performance without delay considerations for the transmitted information. In particular, it considered non-erasure channels and, therefore, no need for feedback.
The seminal work by Ahlswede \textit{et al} \cite{ahslwede00} showed that network coding achieved multicast capacity. In \cite{li03} and \cite{medard03} linear codes over a network were shown to be sufficient to implement any feasible multicast connection, with nodes transmitting a linear combination of the packets previously received. Also, \cite{ho06} showed that linear codes generated randomly in a network achieves multicast capacity. 

For networks where packet erasures exist, two approaches have been used. The first approach relies on rateless codes, i.e. transmitting coded data packets until the receiver sends an acknowledgement stating that all data packets have been decoded successfully. Studies vary from capacity results for random linear network coding \cite{lun08}, trade-off between memory usage and achievable rate for nodes with a fixed, finite memory \cite{lun06wiopt}, network codes that preserve the communication efficiency of a random linear code while achieving better computational efficiency \cite{maymounkov06}, and practical implementations \cite{chachulski07}. Rateless codes have also been studied in terms of delay performance gains and scaling laws(\cite{eryilmaz06}, \cite{ahmed07}, respectively). Finally, queueing analysis has been performed e.g.\cite{jaykumar08}. 

The second approach considers block transmissions. Reference \cite{dana06} considered the problem of wireless networks and showed that linear codes achieve capacity in the network, while Ref. \cite{shrader07} presented a queueing model for random linear coding which codes data packets in a block-by-block fashion using acknowledgements to indicate successful transmission of each block. 

	
	Reference \cite{lucaniInfocom09} considered, for the first time, the use of network coding in channels in which time division duplexing is necessary, i.e. when a node can only transmit or receive, but not both at the same time. This type of channel is usually called half-duplex, but we will use the more general term time division duplexing (TDD) to emphasize the fact that the transmitter and receiver do not use the channel in any pre-determined fashion, but instead may {\em vary} the amount of time allocated to transmit and receive. Important examples of time division duplexing channels are infrared devices (IrDA), and underwater acoustic modems.Other applications may be found in very high latency channels, e.g. in satellite, and deep space communications.  

In particular, Reference \cite{lucaniInfocom09} studied the problem of transmitting $M$ data packets through a link using random linear network coding with the objective of minimizing the expected time to complete transmission of the $M$ data packets. 


Here, we extend the work in \cite{lucaniInfocom09} by focusing on the problem of energy consumption. We show that there exists, under the minimum energy criterion, an optimal number of coded data packets to be transmitted back-to-back before stopping to wait for an acknowledgment (ACK).


By adjusting the number of packets transmitted, our scheme is shown to consume much less energy on average than a full duplex network coding scheme. 
We also show that choosing the number of coded data packets to optimize mean completion time, as in \cite{lucaniInfocom09}, provides a good trade-off between energy consumption and completion time.   
	

	
	 The paper is organized as follows. In Section 2, we outline the problem. In Section 3, we analyze the mean energy needed to complete transmission of $M$ data packets and obtain the optimal transmission algorithm. In Section 4, numerical results are presented to compare the TDD network coding optimized for energy, for completion time, and a full duplex scheme. Conclusions are summarized in Section 5. 

\section{Random Network Coding for TDD channels}

\begin{figure}[t]
\centering	
\includegraphics[height=1in,width=3.5in, keepaspectratio]{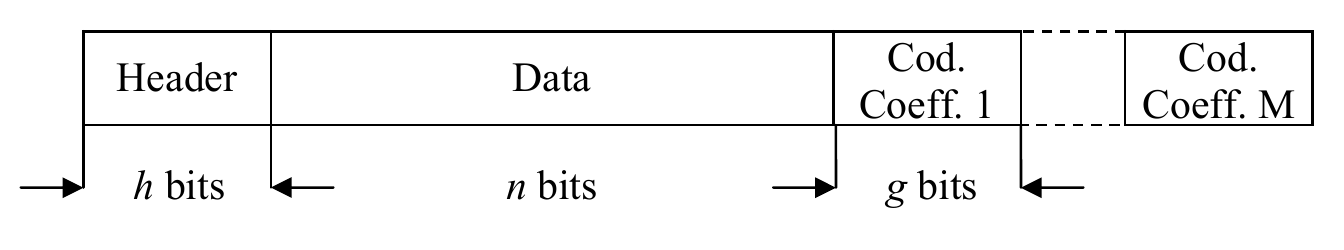}
\caption{\cite{lucaniInfocom09} Structure of coded data packet: a header of size $h$ bits, $n$ data bits, $M$ coding coefficients (Cod. Coeff.) of size $g$~bits each.}
\label{PacketFrame.tag}
\end{figure}

	A sender in a link wants to transmit $M$ data packets at a given link data rate $R$ [bps]. The channel is modeled as a packet erasure channel. Nodes can only transmit or receive, but not both at the same time. The sender uses random linear network coding \cite{ho06} to generate coded data packets. Each coded data packet contains a linear combination of the $M$ data packets of $n$ bits each, as well as the random coding coefficients used in the linear combination. Each coefficient is represented by $g$ bits. For encoding over a field size $q$, we have that $g = \log_2 q$ bits. A coded packet is preceded by an information header of size $h$. Thus, the total number of bits per packet is $h + n + gM$. Figure \ref{PacketFrame.tag} shows the structure of each coded packet considered in our scheme. 

	The sender can transmit coded packets back-to-back before stopping to wait for the ACK packet. The ACK packet feeds back the number of degrees of freedom (dof), that are still required to decode successfully the $M$ data packets. As in \cite{lucaniInfocom09}, we assume that the field size $q$ is large enough so that the expected number of successfully received packets at the receiver, in order to decode the original data packets, is approximately $M$. This is not a necessary assumption for our analysis as discussed in \cite{lucaniInfocom09}; however, it simplifies the expressions and provides a good approximation for large enough $q$, e.g. $q\geq$~1024 which corresponds to $g\geq$~10~bits.


\begin{figure}[t]
\centering	
\includegraphics[height=2.5in,width=2in,keepaspectratio]{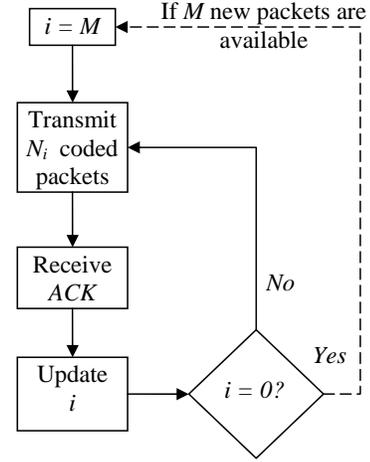}
\caption{\cite{lucaniInfocom09} Algorithm for network coding over time division duplexing channels. $i$ represents the remaining number of degrees of freedom needed at the receiver to decode the packets, and $N_i$ the corresponding number of coded packets that the transmitter will send before stopping to listen for a new ACK. The ACK packet carries the information that will be used to update the value of $i$.}
\label{algorithm.tag}
\end{figure}    	

\begin{figure}[t]
\centering	
\includegraphics[height=1in,width=3.5in]{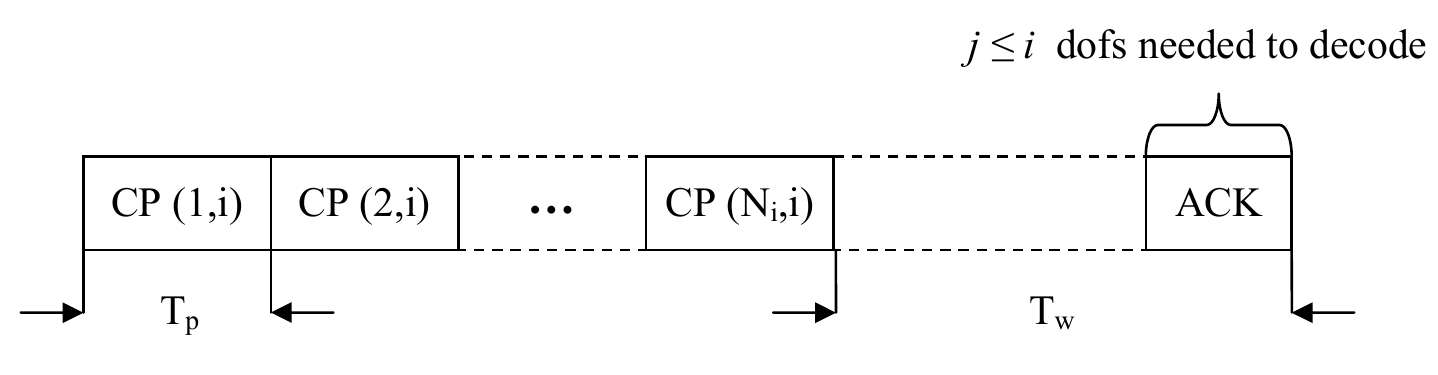}
\caption{\cite{lucaniInfocom09} Network coding TDD scheme.}
\label{Protocol.tag}
\end{figure}

	
Figure \ref{algorithm.tag} illustrates the transmission algorithm. 
Transmission begins with $M$ information packets, which are encoded into $N_M \geq M$ random linear coded packets, and transmitted. If all $M$ packets are decoded successfully, the process is completed. Otherwise, the ACK informs the transmitter how many are missing, say $i$. The transmitter then sends $N_i$ coded packets, and so on, until all $M$ packets have been decoded successfully. We are interested in the optimal number $N_{i}$ of coded packets to be transmitted back-to-back. 

Figure~\ref{Protocol.tag}, illustrates the time window allocated to the system to transmit $N_i$ coded packets. Each coded packet $CP(1,i)$, $CP(2,i)$, etc. is of duration $T_p$. The waiting time $T_w$ is chosen so as to accommodate the propagation delay and time to receive and ACK.

	The process is modelled as a Markov Chain. The states are defined as the number of dofs required at the receiver to decode successfully the $M$ packets. Thus, the states range from $M$ to 0. This is a Markov Chain with $M$ transient states and one recurrent state (state 0). Note that the time and energy spent in each state depends on the state itself, because $N_{i} \neq N_{j}, \forall i \neq j$ in general.       
	
	The transition probabilities from state $i$ to state $j$ ($P_{i\rightarrow j}$) can be defined for any value of $N_i \geq 1$ as follows:
\begin{eqnarray}
P_{i\rightarrow j} = (1-Pe_{ack}) f(i,j) {(1-Pe)}^{i-j} {Pe}^{N_{i} -i + j}
\end{eqnarray}		
where 
\begin{align}
f(i,j) =
\begin{cases}
 \binom{N_{i}}{i - j} & \text{if $N_{i} \geq i$,}
\\
0 & \text{otherwise}
\end{cases}
\end{align}
For $j = i$ the expression for the transition probability is:
\begin{eqnarray}
P_{i\rightarrow i} = (1-Pe_{ack}){Pe}^{N_{i}} + Pe_{ack}
\end{eqnarray}		

\section{Expected Energy for Completing Transmission} \label{SectionExpectedTime.tag}

	The expected energy for completing the transmission of the $M$ data packets is closely related to the expected time of absorption, i.e. the time to reach state 0 for the first time, given that $M$ is the starting state. This can be expressed in terms of the mean energy for completing the transmission given that the Markov Chain is in state is $i$, $E_{i}$ , $\forall i = 0, 1, .. M - 1$. Let us define $E^i$ as the energy consumed by the system to transmit $N_i$ packets and receive an ACK. Then, for $i\geq1$: 
\begin{eqnarray}
E_{i} &=& \scriptstyle \frac{ E^i}{(1-Pe_{ack})(1 - Pe^{N_{i}})}\notag\\ 
&+& \scriptstyle \frac{  {(1 - Pe)}^i {Pe}^{N_{i} - i} \sum _{j = 1} ^{i-1} f(i,j) {\left( \frac{Pe}{1 - Pe} \right)}^j E_{j}}{1 - Pe^{N_{i}}}
\end{eqnarray} 

For the following sections, we consider the case of $E^i = N_{i} E_{p} + E_{ack}$, where $E_{p}$ is the transmission energy of a coded packet, and $E_{ack}$ is the transmission energy of an ACK packet. That is, we consider the case in which transmission energy is dominant in the total energy consumption $E^i$. In other words, the energy used at the receiver and transmitter while waiting for a coded packet and a ACK, respectively, is negligible. However, note that our analysis remains unchanged as long as every $E^i$ is a function solely of $N_i$, which is a reasonable assumption.
More specifically, we define $E_{p} = P T_{p}$, $P$ is the transmission power, $T_{p} = \frac{h +n+gM}{R}$ is the transmission time of a data packet, $E_{ack} = P T_{ack}$, $T_{ack} = n_{ack}/R$ is the transmission time of an ACK, $n_{ack}$ is the number of bits in the ACK packet. Then, we have:
	
\begin{eqnarray}
E_{i} &=& \scriptstyle \frac{ N_{i} E_{p} + E_{ack}}{(1-Pe_{ack})(1 - Pe^{N_{i}} }\notag\\ 
&+& \scriptstyle \frac{  {(1 - Pe)}^i {Pe}^{N_{i} - i} \sum _{j = 1} ^{i-1} f(i,j) {\left( \frac{Pe}{1 - Pe} \right)}^j E_{j}}{1 - Pe^{N_{i}}}\label{MeanEnergy}
\end{eqnarray} 
 
As it can be seen, the mean energy for each state $i$ depends on the mean energies for the previous states. Because of the Markov property, we can optimize the values of all $N_{i}$'s in a recursive fashion, i.e. starting with $N_{1}$, then $N_{2}$ and so on, until $N_{M}$, in order to minimize the expected transmission energy. We do so in the following subsection.      
 
\subsection{Minimizing Expected Energy for Completing Transmission} 

	One objective of this work is to minimize the value of the expected transmission energy $E_{M}$ to transmit $M$ data packets from transmitter to receiver. For $N_{i} \geq 1$, we have:
\begin{eqnarray}
&&\min_{N_{M},..,N_{1}} E_{M} = \min_{N_{M}} \scriptstyle   \frac{  N_{M} E_{p} + E_{ack} }{(1-Pe_{ack})(1 - Pe^{N_{M}})} \normalsize \notag\\
&&+ \scriptstyle \frac{ {(1 - Pe)}^M {Pe}^{N_{M} - M} \sum _{j = 1} ^{M-1} f(M,j) {\left( \frac{Pe}{1 - Pe} \right)}^j  \min_{N_{j},..,N_{1}}  E_{j}}{1 - Pe^{N_{M}}} \normalsize \notag
\end{eqnarray} 
Hence, the problem of minimizing $E_{M}$ in terms of the variables $N_{M},..,N_{1}$ can be solved iteratively. First, we compute $\min_{N_{1}}  E_{1}$, then use this results in the computation of $\min_{N_{2},N_{1}}  E_{2}$, and so on. This characteristic is similar to the iterative characteristic in \cite{lucaniInfocom09} and is related to the Markov structure of the problem.

	One approach to computing the optimal values of $N_{i}$ is to take the derivative of $E_{i}$ with respect to $N_{i}$ treating it as a continuous-valued variable, and look for the value that sets it equal to zero. For our particular problem, this approach leads to solutions without a closed form, i.e. expressed as an implicit function. For $M = 1$, the optimal value of $N_1$ can be expressed using a known implicit function, and it is given by 
\begin{eqnarray}
N_{1}^* = \scriptstyle \frac{ 1 + W \left( - \exp{\left( -1 + \frac{\ln(Pe) E_{ack}}{E_{p}}  \right)} \right)    }{\ln {Pe} }\displaystyle  - \scriptstyle \frac{E_{ack}}{E_{p}} \displaystyle
\end{eqnarray}	
where $W(\cdot)$ is the Lambert W function. The positive values are found for the branch $W_{-1}$, similar to the problem in \cite{lucaniInfocom09}. 
The case of $M = 1$ can be thought as an optimized version of the uncoded Stop-and-Wait ARQ, which is similar to the idea presented in \cite{sastry75}. Instead of transmitting one packet and waiting for the ACK, our analysis suggests that there is an optimal number of back-to-back repetitions of the same data packet that should be transmitted before stopping to listen for an ACK packet. 

 Instead of using the previous approach, we perform a search for the optimal values $N_i, \forall i \in \{1,...M\}$, using  integer values similar to our work in \cite{lucaniInfocom09}. Thus, the optimal $N_i$'s can be computed numerically for given $Pe$, $Pe_{ack}$, $E_{ack}$ and $E_{p}$. In particular, the search method for the optimal value can be made much simpler by exploiting the recursive characteristic of the problem, i.e. instead of making an $M$-dimensional search, we can perform $M$ one-dimensional searches. Finally, the $N_i$'s do not need to be computed in real time. They can be pre-computed and stored in the receiver as look-up tables. This procedure reduces the computational load on the nodes at the time of transmission.
 
\subsection{Comparison Scheme 1: TDD Network Coding With Minimal Completion Time}	

	We use the approach presented in \cite{lucaniInfocom09} to compute $N_i, \forall i \in \{1,...M\}$ which minimizes the mean completion time. These values are used in expression \eqref{MeanEnergy} to determine the mean energy consumed. The expression for mean time to complete transmission can be found in Ref. \cite{lucaniInfocom09}.
 
\subsection{Comparison Scheme 2: Full Duplex Network Coding}	

	This scheme assumes that nodes are capable of receiving and transmitting information simultaneously, and in that sense it is optimal in light of minimal delay. The sender transmits coded packets back-to-back until an ACK packet for correct decoding of all information ($M$ information packets) has been received. This scheme can be modeled as a Markov Chain, where, as before, the states represent the number of dofs received. The energy spent in each state is the same ($E_{p}$). Once the $M$ packets have been decoded, i.e. $M$ dofs have been received, the receiver sends an ACK packets back-to-back, each consuming $E_{ack}$. We define $T_{rt}$ as the round trip time.  
The mean energy to complete the transmission and get and ACK is:
\begin{eqnarray}
E[\text{Energy}] = \scriptstyle \frac{T_{rt}E_{p}}{T_{p}}\displaystyle  + \scriptstyle \frac{T_{rt}E_{ack}}{2 T_{ack}}\displaystyle  + \scriptstyle \frac{M E_{p}}{1 - Pe}\displaystyle + \scriptstyle \frac{E_{ack}}{1-Pe_{ack}}	
\end{eqnarray} 		

The expression for the mean time to complete transmission using this scheme can be found in Ref. \cite{lucaniInfocom09}.

\section{Numerical Examples}
		
	This section provides numerical examples that compare the performance of the different network coding schemes we have discussed so far, namely the two TDD schemes that optimize mean energy consumption (TDD-E) and mean time to complete transmission (TDD-T), and a full duplex scheme. The comparison is carried out in terms of the mean energy and mean time to complete transmission of $M$ data packets under different packet erasure probabilities, with the objective of showing the trade-off between energy and completion time of the different schemes. 
 
\begin{figure}[t]
\centering	
\includegraphics[height=3.4in,width=3.4in,keepaspectratio]{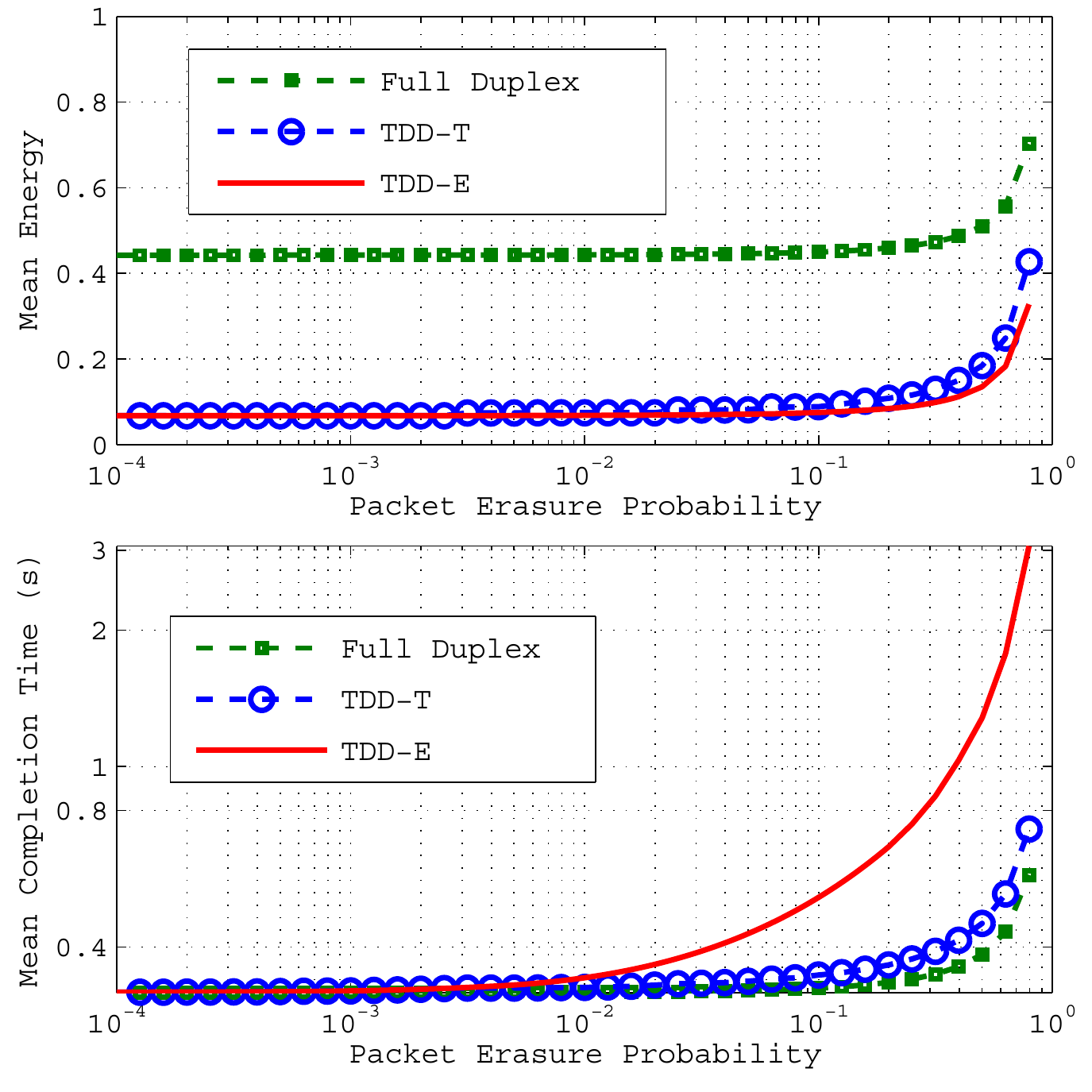}
\caption{Mean Energy and Time to complete transmission. Parameters used: $M=10$, packet size $n = 10,000$ bits, $R=1.5$~Mbps, $h = 80$~bits, $g = 20$~bits, $n_{ack} = 100$~bits.}
\label{ExpectedEnergyLongPropagation.tag}
\end{figure}

Figure \ref{ExpectedEnergyLongPropagation.tag} studies the mean energy and time to complete transmission of $M = 10$ data packets of size $n = 10,000$ bits, with different packet erasure probabilities in a GEO satellite link with a propagation delay of 125~ms. In the following results, we have considered that coded packets and ACK are transmitted with the same power, and that this value is normalized, i.e. $P = 1$. 
 The link parameters are specified in the figure. 

\begin{figure}[t]
\centering	
\includegraphics[height=2.4in,width=3.4in,keepaspectratio]{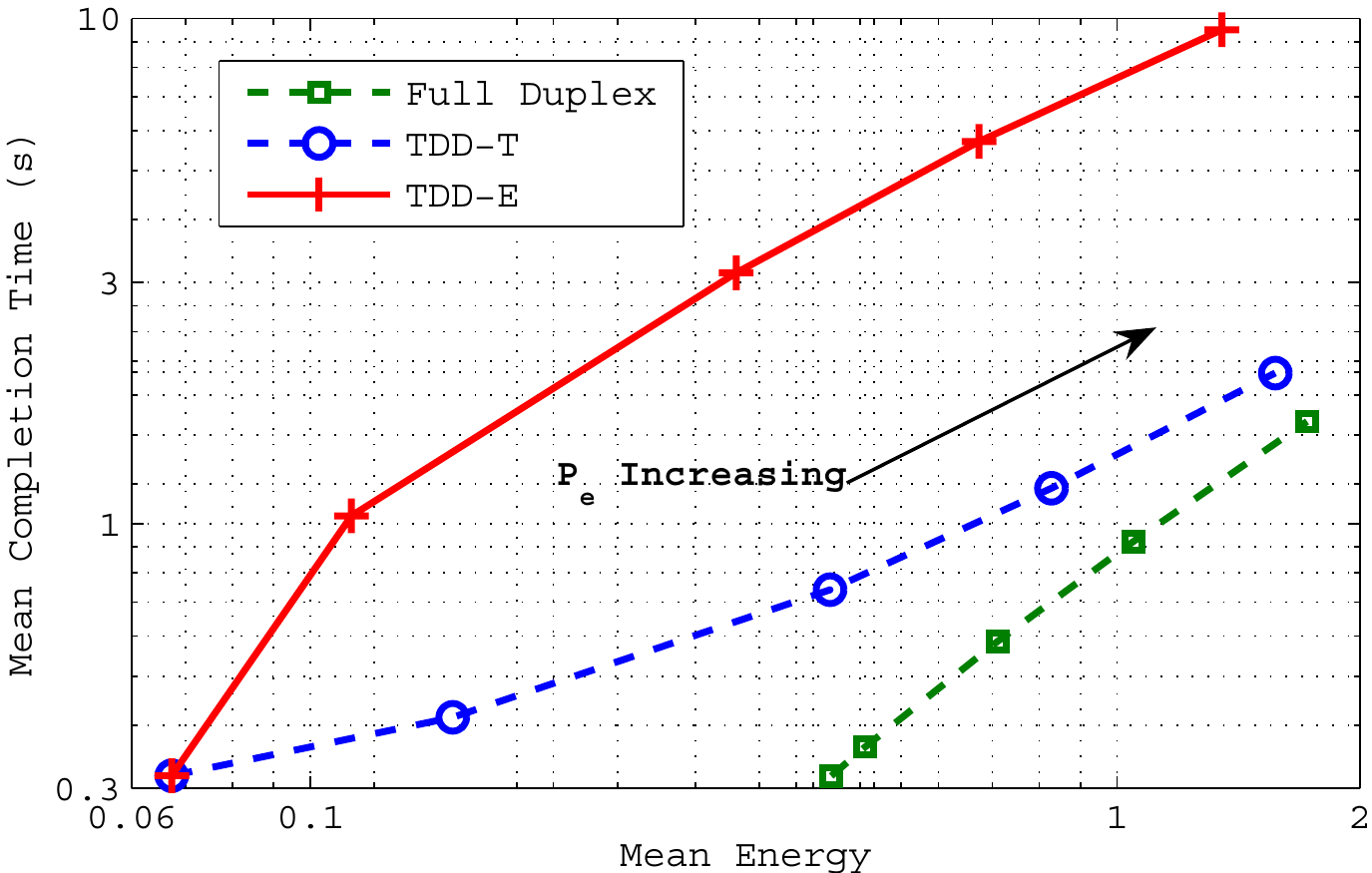}
\caption{Mean Energy and Time to complete transmission trade-off. Parameters used: $M=10$, packet size $n = 10,000$ bits, $R=1.5$~Mbps, $h = 80$~bits, $g = 20$~bits, $n_{ack} = 100$~bits, and $Pe =0.00001, 0.4, 0.8, 0.9, 0.95$.}
\label{TimeEnergyTradeoff.tag}
\end{figure}

The first thing to notice in Figure \ref{ExpectedEnergyLongPropagation.tag} is that both TDD schemes have much better performance with respect to the full duplex scheme, i.e. energy consumption of the full duplex scheme is considerably higher than the TDD schemes given the high latency characteristic of this channel.

 Figure \ref{ExpectedEnergyLongPropagation.tag} shows that the gap between our network coding scheme optimized for energy and for completion time. Their performance stays similar over a wide range of packet erasure probabilities. When the packet erasure probability is low, the performance is the same for the two approaches, both in the sense of energy and delay. For high packet erasure probability the performance of both TDD versions is similar in terms of energy, although we observe a clear advantage of TDD-T over TDD-E in mean completion time.

Figure \ref{ExpectedEnergyLongPropagation.tag} also illustrates that our network coding scheme optimized for completion time (TDD-T) and the network coding full duplex optimal scheme have similar performance over a wide range of packet erasure probabilities. In fact, for the worst case ($Pe = 0.8$) presented in this figure, our scheme has an expected time of completion only 30~\% above the full duplex scheme. Thus, TDD-T can have similar performance to that of full duplex optimal scheme, in the sense of expected time to completion, while showing similar performance to TDD-E, the version optimized for energy consumption. This means that the TDD-T provides a good trade-off between energy and time to complete transmissions.

\begin{figure}[t]
\centering	
\includegraphics[height=3.4in,width=3.4in,keepaspectratio]{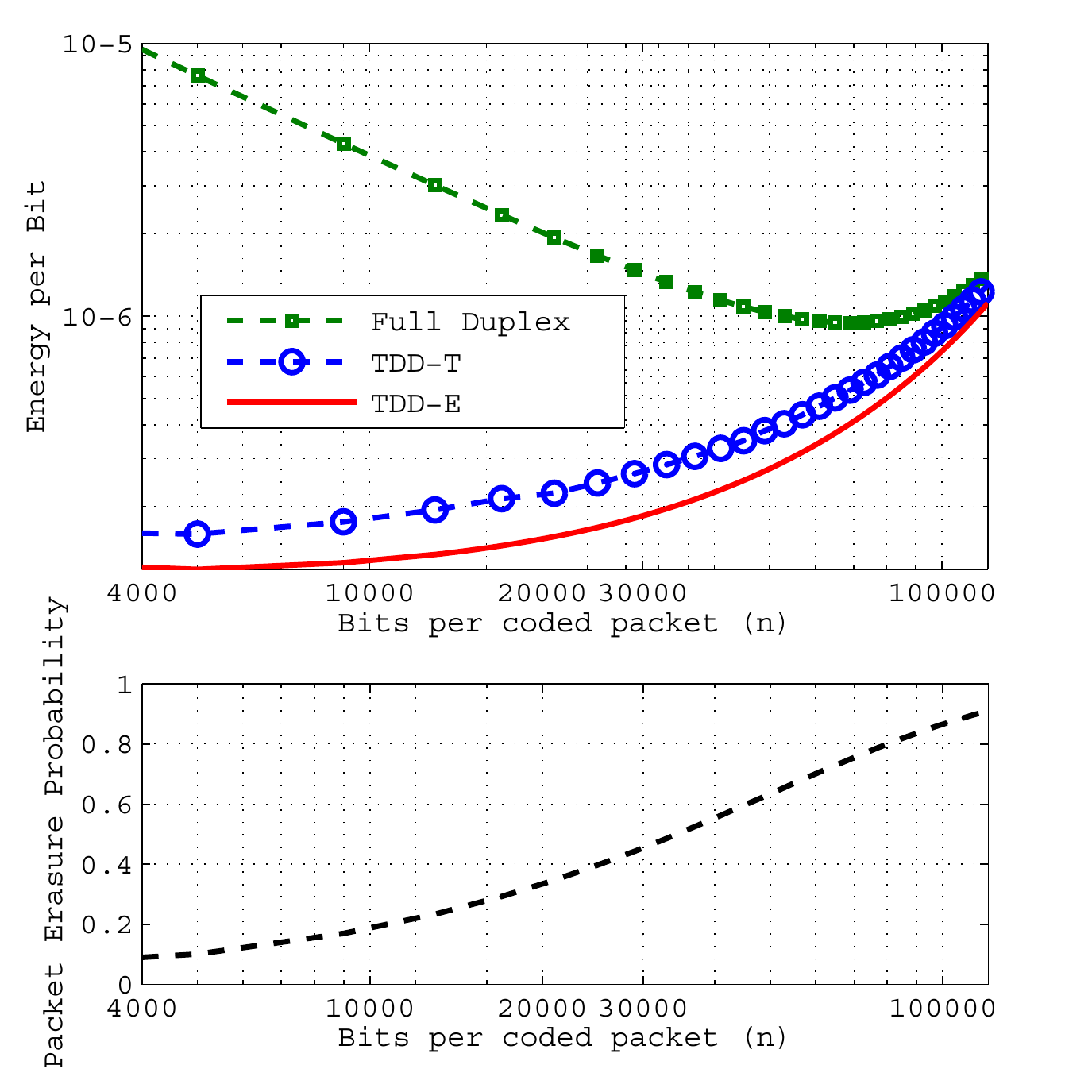}
\caption{Mean Energy per bit to complete transmission with different $n$ bits per coded packet and packet erasure probability for every $n$. Parameters used: $M=10$, $R=10$~Mbps, $h = 80$~bits , $g = 20$~bits, $n_{ack} = 100$~bits, bit error probability is $2.10^{-5}$.}
\label{ExpectedEnergyPerBit.tag}
\end{figure}

Figure \ref{TimeEnergyTradeoff.tag} shows the trade-off curve between time and energy to complete transmission given different packet erasure probabilities $Pe$ ( using same parameters as in Fig. \ref{ExpectedEnergyLongPropagation.tag}). The mean completion time changes significantly between TDD-T and TDD-E as the $Pe$ increases, but the energy difference is small between both schemes. On the other hand, the mean energy is far greater in the full duplex scheme than both TDD schemes, while the completion time of the TDD-T scheme is close to the full duplex scheme.

Figure \ref{ExpectedEnergyPerBit.tag} shows the energy required to successfully transmit one bit of information, i.e. $\frac{E_M}{Mn}$. This Figure considers different values of $n$ in a symmetric channel with parameters in the figure.
It also shows the packet erasure probability for the different values of $n$. First, we notice that the energy per bit required to complete a block transmission is much larger in the full duplex scheme than in both TDD-T and TDD-E. Second, we note that the energy per bit required for both TDD schemes is similar for a wide range of $n$. Finally, the effect of all three schemes having similar energy per bit consumption when $n$ is large is explained by two factors that reduce the effect of the long propagation delay: 1) an increased packet erasure probability which forces more packet transmissions in the TDD schemes, and 2) a larger packet duration $T_{p}$, which causes the number of coded packets in flight to be reduced.
Finally, Figure \ref{ExpectedEnergyPerBit.tag} shows that the full duplex network coding scheme has a value of $n$ that optimizes the energy per data bit required to transmit the entire information.     

\section{Conclusion}
This paper studies the use of random linear network coding over channels where time division duplexing is necessary. In particular, we study the energy consumption in a randomly coded system, and propose a transmission scheme that minimizes the mean energy spent for transmitting a pre-determined number of packets. In this scheme, a number of coded data packets are transmitted back-to-back before stopping to wait for the receiver to acknowledge how many degrees of freedom, if any, are required to decode the information correctly. 
	The optimal number of coded data packets, in terms of mean energy required to complete transmission depends of probabilities of packet and ACK erasure, the energy required to transmit a coded packet and an ACK, and the number of degrees of freedom that the receiver requires to decode the data. Although, there is no closed form solution for the optimal number of packets, a simple search method can be obtained by exploiting the recursive characteristic of the problem, i.e. instead of making an $M$-dimensional search, we perform $M$ one-dimensional searches. 
	
	
	 	We present an analysis and numerical results that show that transmitting the optimal number of coded data packets sent before stopping to listen for an ACK (in terms of both mean completion time and mean energy consumed) consumes much less energy in average than a network coding scheme operating in a full duplex channel to complete transmission of $M$ packets. We also show that choosing the number of coded
data packets to optimize mean completion time, as in \cite{lucaniInfocom09},
provides a good trade-off between energy consumption and completion time.   
	 
	Future research will consider an extension of the principles proposed for one link to the general problem of wireless networks, possible due to the use of random network coding. In this extension, each node transmitting through a link, or, more generally, a hyperarc (using the terminology in \cite{lun06}) will have an optimal number of coded packets to transmit.


\section*{Acknowledgment}
This work was supported in part by the National Science Foundation under grants \# 0520075 and CNS-0627021, by ONR MURI Grant No. N00014-07-1-0738, and subcontract \# 060786 issued by BAE Systems National Security
Solutions, Inc. and supported by the Defense Advanced Research Projects
Agency (DARPA) and the Space and Naval Warfare System Center (SPAWARSYSCEN),
San Diego under Contract No. N66001-06-C-2020.



%

\end{document}